\documentclass{article}
\usepackage{stywhispers,amsmath,epsfig}
\usepackage{amssymb}
\usepackage{graphicx}
\usepackage{subcaption}

\usepackage{bbold}
\usepackage{multirow}
\usepackage{dcolumn} 
\usepackage{booktabs}

\usepackage{algorithm}
\usepackage{algpseudocode}
\newcommand{\NoIndentState}[1]{\State \parbox[t]{\dimexpr\linewidth-\algorithmicindent}{#1\strut}}

\newcolumntype{d}[1]{D{.}{.}{#1}}

\DeclareMathOperator*{\argmin}{arg\,min}

\title{Optimal transport-based hyperspectral unmixing for highly mixed observations}
\name{Delphine Doutsas$^{1}$, Bruno Figliuzzi$^{1}$ \thanks{\noindent1. The authors thank The Transition Institute 1.5 - Mines Paris - PSL University for supporting this work. \\ This work has been accepted and published at the IEEE WHISPER Workshop 2024. © IEEE. Personal use of this material is permitted. Permission from IEEE must be obtained for all other uses.}}
\address{Centre for Mathematical Morphology (CMM), Mines Paris, PSL University}
\begin{document}
%

\maketitle
\begin{abstract}

We propose a novel approach based on optimal transport (OT) for tackling the problem of highly mixed data in blind hyperspectral unmixing. Our method constrains the distribution of the estimated abundance matrix to resemble a targeted Dirichlet distribution more closely. The novelty lies in using OT to measure the discrepancy between the targeted and true abundance distributions, which we incorporate as a regularization term in our optimization problem. We demonstrate the efficiency of our method through a case study involving an unsupervised deep learning approach. Our experiments show that the proposed approach allows for a better estimation of the endmembers in the presence of highly mixed data, while displaying robustness to the choice of target abundance distribution.
\end{abstract}


\section{Introduction}
\label{sec:intro}

Hyperspectral imaging is a specialized acquisition technique that captures images across a wide range of narrow wavelength bands. 
Because of the limited resolution of the image, a single pixel may contain multiple materials or \textit{endmembers}, each contributing to the pixel's spectrum in a certain proportion or \textit{abundance}~\cite{boucas-dias-review}. \newline

\noindent
In blind hyperspectral unmixing (BHU), the goal is to jointly estimate the endmembers and the corresponding abundances at each pixel. Without any further assumption, an infinity of solutions exists, making the problem ill-posed. A common approach is to model the problem using a linear mixing model (LMM) and optimize the Nonnegative Matrix Factorization (NMF) objective function subject to the sum-to-one constraint on the abundance vector. Given the under-determined nature of the problem, it is necessary to introduce stronger inductive biases beyond nonnegativity, either on the endmember matrix or the abundance matrix. These biases can be statistical, as in methods like DECA \cite{DECA}, supposing that the abundances follow a Dirichlet distribution, or geometrical, based on the simplex shaped by the data, such as NFINDR \cite{nfindr} and VCA \cite{vca}. 
Notably, unsupervised deep learning (DL) approaches have become increasingly important. Many architectures propose to learn to map the data into a subspace through auto-encoders \cite{palsson-ae, palsson-cnn-ae}. \newline

\noindent 
In the following, we refer to the situation where only a few or even no pixel spectra lie on the facets of the endmembers simplex as the \textit{highly mixed scenario}. In this scenario, traditional methods fail or deteriorate \cite{boucas-dias-review, figliuzzi, zenati, zenati2022}. Several DL methods attempt to tackle this problem by assuming that the abundances follow a Dirichlet distribution and learn the distribution parameters to sample the abundance vectors~\cite{ldvae, ildvae}. Related work such as the Sinkhorn autoencoder \cite{sinkhorn-ae} has also leveraged optimal transport regularization to learn meaningful latent representations, highlighting the potential of OT-based approaches in high-dimensional mixture modeling. To address this challenge, we propose a novel approach that integrates a regularization term based on optimal transport - the Sinkhorn divergence - applied to the abundance vector distribution. This term combines statistical and geometrical biases in the distribution space to better handle the complexity of highly mixed data. To our knowledge, this is the first time such an optimal transport (OT) regularization term has been applied in blind hyperspectral unmixing. This paper demonstrates how our method enhances the unmixing process by effectively addressing the challenges of highly mixed data, thereby improving the results' accuracy and convergence robustness.


\section{Proposed method}
\label{sec:method}

\subsection{The Linear Mixing Model (LMM)}
In what follows, we have access to observations $\mathbf{Y} = [\mathbf{y}_1, \dots, \mathbf{y}_n] \in \mathbb{R}_+^{l \times n}$, each representing a pixel $\mathbf{y}_i \in \mathbb{R}_+^l$ in the image with spectral dimension $l$. Each observation is assumed to be a linear mixture of $k$ endmembers, concatenated in a matrix $\mathbf{M} \in \mathbb{R}_+^{l \times k}$, with corresponding abundances $\mathbf{A} = [\mathbf{a}_1, \dots, \mathbf{a}_n] \in \mathbb{R}_+^{k \times n}$ satisfying the sum-to-one constraint $\forall i \in \{ 1, \dots, n\}, \; \mathbf{1}_k ^T \mathbf{a}_i  = 1.$
In matrix form, the Linear Mixing Model (LMM) is described by the equation
\begin{equation} 
    \mathbf{Y} = \mathbf{M}\mathbf{A} +  \mathbf{E},
\end{equation}
where the endmember and abundance matrices $\mathbf{M}$ and $\mathbf{A}$ are nonnegative and $E$ is a Gaussian noise matrix.
 
\subsection{Nonnegative Matrix Factorization (NMF)}
In practice, inverting the LMM amounts to addressing the following NMF problem:
\begin{equation} \label{eq:LMM-simple}
\hat{\mathbf{M}}, \hat{\mathbf{A}} = 
\argmin_{%
      \substack{%
        \mathbf{M} \in \mathbb{R}_+^{l \times k} \\
        \mathbf{A} \in \mathbb{R}_+^{k \times n}
      }
    }
    \mathcal{L}_{rec}(\mathbf{Y}, \mathbf{M}, \mathbf{A})
\end{equation} where $\mathcal{L}_{rec}$ is a data-fidelity term. 
Since this problem is ill-posed, regularization is required. This usually leads to a reformulation so that
\begin{equation}  \label{eq:complete-nmf}
\hat{\mathbf{M}}, \hat{\mathbf{A}} = \argmin_{%
\substack{%
\mathbf{M} \in \mathbb{R}^{l \times k} \\ 
\mathbf{A} \in \mathbb{R}^{k \times n} 
} 
} \mathcal{L}_{rec}(\mathbf{Y}, \mathbf{M}, \mathbf{A}) + \lambda_1 f(\mathbf{A}) + \lambda_2 g(\mathbf{M})
\end{equation} where $f$ and $g$ are regularization terms on $\mathbf{A}$ and $\mathbf{M}$, respectively. 
In this paper, $\mathcal{L}_{rec}$ can be indifferently the Mean Squared Error (MSE), 
\vspace{-0.1cm}
$$
\mathcal{L}_\text{MSE}(\mathbf{Y}, \hat{\mathbf{Y}} := \hat{\mathbf{M}}\hat{\mathbf{A}}) := \frac{1}{n} \sum_{i=1}^{n} \left( \hat{\mathbf{y}}_i - \mathbf{y}_i \right)^2, 
$$
\vspace{-0.1cm}
or the Spectral Angle Distance (SAD),
\vspace{-0.1cm}
$$
\mathcal{L}_\text{SAD}(\mathbf{Y}, \hat{\mathbf{Y}}) = \frac{1}{n} \sum_{i=1}^{n} \text{arccos} \left( \frac{\langle \hat{\mathbf{y}}_i, \mathbf{y}_i \rangle }{ \| \hat {\mathbf{y}}_i\|_2 \| \mathbf{y}_i\|_2 } \right).
$$

\vspace{-0.4cm}

\subsection{Auto-Encoder}
As a hands-on case, we focus on methods that employ auto-encoders (AE) to estimate $\mathbf{A}$ and $\mathbf{M}$. An AE consists of two parametric functions, an encoder, and a decoder, combined in the form $\mathcal{G} := \mathcal{G}_D \circ \mathcal{G}_E$. 

AE-based approaches are commonly employed in BHU following the seminal work of~\cite{palsson-ae}. In this work, we adopt their proposed architecture. In their setting, the encoder network $\mathcal G_E$ maps the original observation $\mathbf{y}$ into a latent variable $\mathbf{\hat a} \in \mathbb{R}^k$ interpreted as an estimation of the abundances associated with this observation. We denote by $\mathbf{W}_E$ the weights of the encoder network, which comprises three dense layers, followed by a batch normalization layer and specific layers ~\cite{palsson-ae} ensuring that both non-negativity and sum-to-one constraints are satisfied for $\mathbf{\hat a}$. To match the LMM formulation, the decoder takes the simple form $\mathcal{G}_D(\mathbf{\hat A}) = \mathbf{W}_D\mathbf{\hat A}$ and is trained to reconstruct the original observations matrix $\mathbf{Y}$. It is clear from the equation form (\ref{eq:LMM-simple}) that $\mathbf{W}_D$ can be interpreted as an estimation $\hat{\mathbf{M}}$ of the endmember matrix $\mathbf{M}$. 

\subsection{Optimal Transport Regularization}

In this article, our main contribution is to propose a novel approach based on optimal transport to regularize the NMF formulation of the LMM. Most existing approaches regularize the problem by attempting to minimize the volume of the endmembers simplex. By contrast, we apply the regularization to the abundances to select a solution where they are close to a target distribution, typically chosen as a Dirichlet distribution $\mathcal{D}(\alpha)$ on the $k$-simplex with manually specified parameters $\alpha$. In this setting, a major challenge is to define a distance between the target distribution and the empirical distribution associated with the abundances estimated by the algorithm. In practice, the choice of the Sinkhorn divergence, introduced in \cite{genevay}, is particularly well-suited. In mathematical terms, the NMF problem can thus be defined as
\begin{equation}
    \begin{aligned}
        \mathbf{W}^*_E, \mathbf{W}^*_D = 
     \argmin_{%
      \substack{%
        \mathbf{W}_E, \mathbf{W}_D
      }
    } & \;\;\; \mathcal{L}_{rec}(\mathbf{Y}, \mathcal{G}(\mathbf{Y}; \mathbf{W}_E, \mathbf{W}_D)) \\
      & + \lambda_{reg} \cdot \mathcal{S}_{\varepsilon}(\mathcal D_{emp}(\hat {\mathbf{A}}), \mathcal{D}(\alpha)) 
    \end{aligned}
\end{equation}
where $\hat {\mathbf{A}} := \mathcal{G}_E(\mathbf{Y}; \mathbf{W}_E)$ corresponds to the estimated abundance matrix, $\mathcal S_{\varepsilon}$ denotes the Sinkhorn divergence, $\mathcal D_{emp}(\hat {\mathbf{A}}) := \frac{1}{n} \sum_{i=1}^n \delta_{\mathbf{a}_i}$ the empirical distribution of the estimated abundance vectors, and $\lambda_{reg} \in \mathbb R$ the regularization factor. The weights $\mathbf{W}^*_E$ provide an estimation of the endmember matrix. 
\newline

\noindent
\textbf{Sinkhorn divergence} 
Suppose we have two finite sets of points \(\mathbf{X} = \{\mathbf{x}_1, \dots, \mathbf{x}_m\}\) and \(\mathbf{Y} = \{\mathbf{y}_1, \dots, \mathbf{y}_{m'}\}\), and two discrete probability measures $\mu = \sum_{i=1}^{m} \mu_i \delta_{\mathbf{x}_i}$ and  $\nu = \sum_{j=1}^{m'} \nu_j \delta_{\mathbf{y}_j}$ where \(\delta_{\mathbf{x}_i}\) and \(\delta_{\mathbf{y}_j}\) are Dirac masses. Let \(c(\mathbf{x}_i, \mathbf{y}_j) = \|\mathbf{x}_i - \mathbf{y}_j\|_2\) be the (Euclidean) cost of transporting unit mass from \(\mathbf{x}_i\) to \(\mathbf{y}_j\). We want to devise an optimal transport plan $\gamma$ that would "move" the mass described by $\mu$ to the mass of $\nu$ while minimizing the distance to be covered.
In this discrete case, the transport plan $\gamma \in \mathbb{R}^{m \times m'}$ is a matrix, where each element $\gamma_{ij} \geq 0$ represents the amount of mass transported from $\mathbf{x}_i$ to $\mathbf{y}_j$. In particular, $\gamma$ is a discrete distribution defined on $X \times Y$ that must satisfy constraints (\ref{eq:marginal-constraints}) on its marginals.
\vspace{-0.2cm}
\begin{equation} \label{eq:marginal-constraints}
    \forall i, \quad \sum_{j=1}^{m'} \gamma_{ij} = \mu_i \quad \text{ and } \quad \forall j, \sum_{i=1}^{m} \gamma_{ij} = \nu_j.
\end{equation}
We denote by $\Gamma(\mu, \nu)$ the set of all discrete distributions on  $X \times Y$ satisfying these constraints. The OT problem takes the form of an optimization problem (\ref{eq:optimal-transport-plan}) in which the optimal $\gamma^*$ provides the minimal transport cost $C$,
\vspace{-0.4cm}
\begin{equation} \label{eq:optimal-transport-plan}
C^* = \inf\limits_{\gamma \in \Gamma(\mu , \nu)} C(\gamma ) := \sum_{i=1}^{m} \sum_{j=1}^{m'} \gamma_{ij} \|\mathbf{x}_i -  \mathbf{y}_j \|_2.
\end{equation}
This minimal cost defines a distance between the empirical distributions, referred to as the \textit{Wasserstein distance} in the literature. 

To evaluate this distance, one must solve (\ref{eq:optimal-transport-plan}), which is computationally demanding. To facilitate the resolution of the problem, it is common to add an entropic regularization term based on the KL divergence. This yields the so-called $\varepsilon$-Wasserstein distance, defined as
\begin{equation} \label{eq:entropic-wasserstein-distance}
    \begin{aligned}
        \mathcal W_{\varepsilon}(\mu, \nu) := \min\limits_{\gamma \in \Gamma(\mu, \nu)} C(\gamma ) +\varepsilon \cdot KL (\gamma \mid \mu \otimes \nu)
    \end{aligned}
\end{equation}
where $\varepsilon \in \mathbb R$ is the entropic factor. As demonstrated in \cite{marco-cuturi-sinkhorn}, problem (\ref{eq:entropic-wasserstein-distance}) can be efficiently solved using the Sinkhorn algorithm. A valuable feature of this algorithm is that it allows for the efficient computation of the $\varepsilon$-Wasserstein distance and its gradient with respect to the abundance matrix.

\noindent
It can easily be noted that the $\varepsilon$-Wasserstein distance is not a distance in the strict sense, because $\mathcal W_{\varepsilon}(\mu, \mu) \neq 0$. In \cite{genevay}, it is therefore proposed to replace this quantity with the so-called Sinkhorn divergence defined by
\begin{equation}
\mathcal{S}_{\varepsilon}(\mu, \nu ) = 2\mathcal W_{\varepsilon}(\mu, \nu) -  \mathcal W_{\varepsilon}(\mu, \mu) - \mathcal W_{\varepsilon}(\nu, \nu). 
\end{equation}

\vspace{0.2cm}

The regularization term involved in the formulation (\ref{eq:complete-nmf}) of the NMF problem consists in computing the Sinkhorn divergence between the empirical distribution $\mathcal D_{emp}(\hat A)$ associated with the estimated abundances and a continuous distribution $\mathcal{D}(\alpha )$. To that end, we can compute an approximation of this term by replacing $\mathcal{D}(\alpha )$ with the empirical distribution $\hat{\mathcal{D}}(\alpha ) = \frac{1}{m'}\sum_{i = 1}^{m'} \delta _{\mathbf{x}_i}$ where $(\mathbf{x}_i)_{i = 1, \dots m'}$ are i.i.d samples from $\mathcal{D}(\alpha )$ and use Sinkhorn algorithm.
\newline

\noindent
\textbf{Algorithm} Our proposed algorithm is summarized below. 
\begin{algorithm} 
\caption{Proposed algorithm}
\label{alg:update} 
\begin{algorithmic}[1] 
    \State \textbf{Input:} Endmember and abundance matrices $\mathbf{\hat{M}}_0$, $\mathbf{\hat{A}}_0$, number of epochs $T$, batch size $m$, $\lambda_{reg}$, learning rate $\eta$, momentum $\tau$.
    
        \For{$t = 0$ to $T$} 
        
            \For{each batch} 

                \NoIndentState{Compute  $\hat{\mathbf Y} = \mathcal{G}_D \circ \mathcal{G}_E(\mathbf{Y})$}
                \State{$\mathbf{\hat{A}}_t \leftarrow \mathcal{G}_E(\mathbf{Y})$}
                \State{$\mathbf{\hat{M}}_t \leftarrow \mathbf{W}_D$} 
            
                \NoIndentState{Compute $ \mathcal{L}_{\text{rec}}(\mathbf{Y}, \mathbf{\hat{M}}_t, \mathbf{\hat{A}}_t) = \| \mathbf{Y} - \mathbf{\hat{M}}_t \mathbf{\hat{A}}_t \|^2 $}
 
                \NoIndentState{Sample $m'$ points $X := (\mathbf{x}_i)_{i=1}^{m'}$ from $\mathcal{D}(\alpha)$} 
                
                \NoIndentState{Compute 
                \vspace{-0.4cm}
                $$ \mathcal{L}_{reg}(\mathbf{\hat{A}}_t, X) = \mathcal{S}_{\varepsilon} \left( \frac{1}{m} \sum_{i=1}^{m} \delta_{\mathbf{\hat{a}}_i}, \frac{1}{m'} \sum_{i=1}^{m'} \delta_{\boldsymbol{x}_i} \right) $$
                }
                \vspace{-0.7cm}

                \NoIndentState{Compute 
                $\mathcal{L}_{tot} = \mathcal{L}_{\text{rec}} + \lambda_{reg} \cdot \mathcal{L}_{reg}$}
                
                \NoIndentState{Compute $\nabla_{\mathbf{\hat{M}}, \mathbf{\hat{A}}} \mathcal{L}_{\text{tot}}$ using backpropagation}
                
                \NoIndentState{ Update AE weights with RMSprop}
                                
            \EndFor
            
        \EndFor 
\end{algorithmic} 
\end{algorithm}
 
\vspace{-2mm}

\section{Experiments / Results}
\label{sec:results}

\subsection{Experimental Settings}

\textbf{Data settings} Our proposed algorithm is evaluated on a synthetic dataset referred to as \textit{Synth4} in the remainder of the article, and on a real dataset, \textit{Samson}~\cite{zhu}. 
\textit{Synth4} contains $2000$ samples and is generated using 3 endmember spectra with 224 spectral bands randomly selected in the USGS Spectral Library \cite{usgs-library}. The abundances are simulated by sampling from a Dirichlet distribution with concentration parameters $\alpha = (4, 4, 4)$ to generate highly mixed spectra. An SNR of 40 is assumed to simulate realistic observations, and some Gaussian noise is introduced accordingly. In addition, the maximum purity is fixed to $0.8$. 
\textit{Samson} contains 9025 samples. Our implementations is based on the cleaned version of \textit{Samson} containing 156 bands after denoising with Principal Component Analysis projection (PCA). The ground truth endmembers are given normalized while the observations are not.

The main characteristic of \textit{Synth4} dataset is the highly mixed observations. Indeed, the abundances are sampled according to a Dirichlet distribution with parameters $(4, 4, 4)$, which yields spectral observations that are concentrated around the center of the endmembers simplex. In particular, the dataset contains neither pure pixels nor pixels on the facets of the simplex. For this dataset, it is therefore expected that the introduction of the regularization term based on optimal transport will lead to a significant improvement in the endmembers estimation. By contrast, the nature of the \textit{Samson} dataset is very different. First, it must be noted that \textit{Samson} observations are not highly mixed: the dataset contains for instance a significant number of pure pixels for each endmember. Second, for this dataset, the exact distribution of the abundances is not known \textit{a priori}. Therefore, we select a uniform Dirichlet distribution on the simplex for the regularization term. In practice, this Dirichlet distribution does not correspond to the real distribution of the abundances, and the main objective of our experiment is therefore to verify that the method proposed in this article remains robust to the  prior choice of the target abundance distribution.\newline

\noindent
\textbf{Training settings}
The concentration set of the target Dirichlet distribution is set manually to $\alpha = (4, 4, 4)$ (resp. $(1, 1, 1)$) for the dataset \textit{Synth4} (resp. \textit{Samson}). For \textit{Samson}, as mentioned previously, it is important to note that this choice significantly differs from the true abundance distribution. 
The computation of the Sinkhorn divergence depends on selecting an entropic factor that balances the trade-off between the quality of the optimal plan and the computational efficiency. In this work, a value of $\varepsilon=10^{-2}$ is found to effectively satisfy both criteria.
\noindent
Our autoencoder is adapted from the TensorFlow implementation in \cite{palsson-ae} to PyTorch, with the sum-to-one and sparse-ReLU layers re-implemented. Xavier-uniform initialization replicates TensorFlow's default settings for the encoder's dense layers, as in \cite{palsson-ae}. However, the decoder's dense layer is initialized using NFINDR applied to the training dataset \cite{nfindr}, which differs from the original approach in \cite{palsson-ae}.
As the goal is to balance the training curves without comparing the predicted endmembers to the ground truth, the reconstruction and optimal transport losses are the only available information to guide the hyperparameters' choice. In practice, a learning rate set to $10^{-2}$ and a regularization factor $\lambda_{reg}=10$ fulfill this adjusting purpose for both \textit{Synth4} and \textit{Samson}. 
Finally, $200$ samples are drawn from the continuous target Dirichlet distribution at each batch in an epoch during training to estimate the OT-based regularization term. The batch size is randomly chosen to be 185. 

\subsection{Results}

%

\begin{figure}[t]
    \begin{subfigure}[t]{0.5\textwidth}
        \centering
        \includegraphics[width=8.5cm]{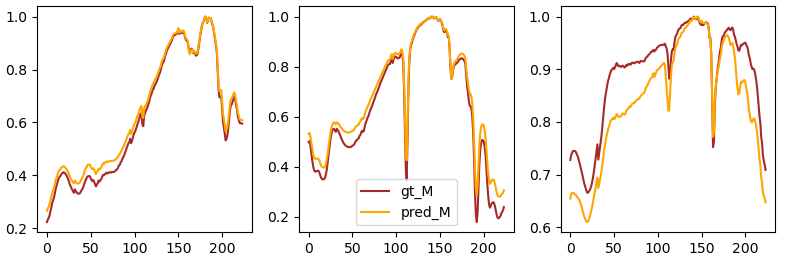}
        \caption{AE with OT regularization.}
    \end{subfigure}%
    \hfill
    \begin{subfigure}[t]{0.5\textwidth}
        \centering
        \includegraphics[width=8.5cm]{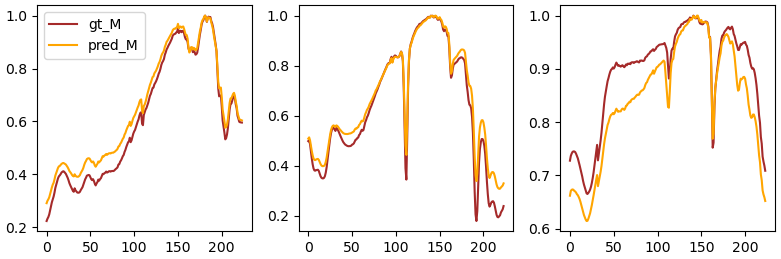}
        \caption{AE without OT regularization.}
    \end{subfigure}
    \caption{Example of predicted endmembers by AE with and without OT on \textit{Synth4} dataset, at epoch 20.}
    \label{fig:444-predictions-of-aes}
\end{figure}

\begin{figure}[t]
    \begin{subfigure}[t]{0.5\textwidth}
        \centering
        \includegraphics[width=8.5cm]{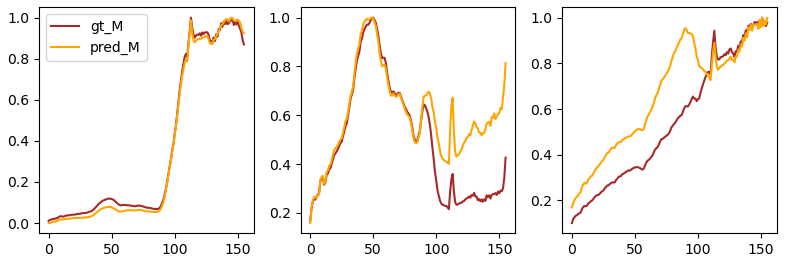}
        \caption{AE with OT regularization.}
    \end{subfigure}%
    \hfill
    \begin{subfigure}[t]{0.5\textwidth}
        \centering
        \includegraphics[width=8.5cm]{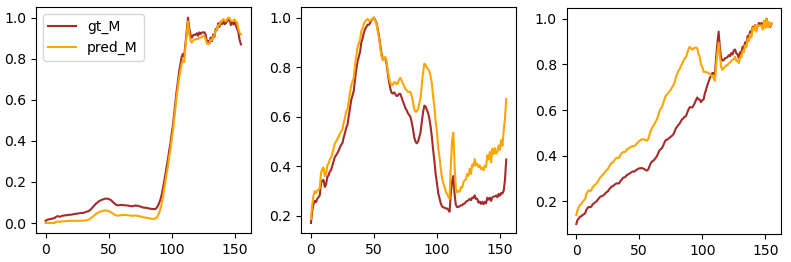}
        \caption{AE without OT regularization.}
    \end{subfigure}
    \caption{Example of predicted endmembers by AE with and without OT on Samson dataset, at epoch 20. From left to right: tree, water, soil}
    \label{fig:samson-predictions-of-aes}
\end{figure}

Tables~\ref{tab:acc_on_ems_sad_444} and~\ref{tab:acc_on_ems_sad_samson} compare the abundance estimation performance of an AE trained with different objective functions, with and without OT regularization for \textit{Synth4} and \textit{Samson} datasets. In the absence of the regularization term, the algorithm replicates the recent approach proposed by Palsson et al. in~\cite{palsson-ae}. Figures \ref{fig:444-predictions-of-aes} and \ref{fig:samson-predictions-of-aes} display the predicted endmembers for an AEs trained on 20 epochs with an MSE objective function, with (\ref{fig:444-predictions-of-aes}a) and without (\ref{fig:samson-predictions-of-aes}b) OT regularization for \textit{Synth4} and \textit{Samson}.


\begin{table}[htbp!] 
    \centering
    \resizebox{0.48\textwidth}{!}{%
    \begin{tabular}{l p{1.95cm} p{1.95cm} p{1.95cm} p{1.95cm}}
        \toprule
         & \multicolumn{1}{c}{\textbf{SAD}~\cite{palsson-ae}} & \multicolumn{1}{c}{\textbf{OT+SAD}} & \multicolumn{1}{c}{\textbf{MSE}~\cite{palsson-ae}} & \multicolumn{1}{c}{\textbf{OT+MSE}} \\
        \midrule
        Em 1 & 0.110 ± 0.017 & 0.090 ± 0.039 & 0.164 ± 0.015 & 0.083 ± 0.030 \\
        Em 2 & 0.074 ± 0.002 & 0.072 ± 0.026 & 0.210 ± 0.018 & 0.058 ± 0.022 \\
        Em 3 & 0.095 ± 0.024 & 0.091 ± 0.019 & 0.046 ± 0.024 & 0.062 ± 0.030 \\
        \bottomrule
    \end{tabular}
    }
    \caption{Endmember estimation results for \textit{Synth4} dataset. The values are obtained on the 5 best runs with the SAD as a metric.  The name of each column specifies the objective loss(es) used during training. \label{tab:acc_on_ems_sad_444}}
\end{table}

\begin{table}[htbp!] 
    \centering
    \resizebox{0.48\textwidth}{!}{%
    \begin{tabular}{l p{1.95cm} p{1.95cm} p{1.95cm} p{1.95cm}}
        \toprule
         & \multicolumn{1}{c}{\textbf{SAD}~\cite{palsson-ae}} & \multicolumn{1}{c}{\textbf{OT+SAD}} & \multicolumn{1}{c}{\textbf{MSE}~\cite{palsson-ae}} & \multicolumn{1}{c}{\textbf{OT+MSE}} \\
        \midrule
        Soil  & 0.071 ± 0.006  & 0.086 ± 0.037 & 0.124 ± 0.062 & 0.281 ± 0.071  \\
        Tree  & 0.049 ± 0.008  & 0.082 ± 0.004 & 0.061 ± 0.011 & 0.162 ± 0.084 \\
        Water & 0.201 ± 0.022  & 0.269 ± 0.004 & 0.251 ± 0.073 & 0.255 ± 0.084 \\
        \bottomrule
    \end{tabular}
    }
    \caption{Endmember estimation results for \textit{Samson} dataset. The values are obtained on the 5 best runs with the SAD as a metric. The name of each column specifies the objective loss(es) used during training. \label{tab:acc_on_ems_sad_samson}}
\end{table}





From Table \ref{tab:acc_on_ems_sad_444}, it can be observed that our method exhibits better performances in the highly mixed scenario (\textit{Synth4}) when compared to~\cite{palsson-ae}. This performance improvement is due to using a regularization term based on optimal transport, which forces the selection of endmembers such that the resulting abundances effectively follow a Dirichlet distribution with parameters $(4, 4, 4)$. Introducing the regularization term thus significantly improves the estimation of the endmembers.
Our approach shows however lower performances on the \textit{Samson} dataset than the traditional approach. We attribute this to the consequence of selecting a target Dirichlet distribution that differs from the actual abundance distribution and to the fact that \textit{Samson} contains pure pixels for each endmember, reducing the need for OT regularization. Nevertheless, Fig. \ref{fig:samson-predictions-of-aes} indicates that even if the SAD metrics are lower when using the OT regularization, our approach exhibits significant robustness to the choice of the target abundance distribution and remains able to estimate the endmembers properly. The experiment conducted on \textit{Samson} dataset therefore illustrates that the OT regularization approach remains robust when applied in situations where the target distribution diverges from the actual abundance distribution.



\section{Conclusion}

In this article, we presented a novel approach for performing BHU on highly mixed observations, relying on optimal transport to regularize the abundances assuming they follow a predefined target distribution.
Our experiments demonstrated that the proposed approach allows for better estimation of the endmembers in the presence of highly mixed data, compared to recent BHU approaches \cite{palsson-ae}.

A limitation of our method is that the target abundance distribution used for OT regularization must be manually defined. Nevertheless, experiments conducted on the \textit{Samson} dataset demonstrated that our approach remains robust to the choice of the target abundance distribution. Selecting an appropriate target distribution for the abundance is a natural extension of this study, which will be the focus of future work.


\label{sec:conclusion}


\end{document}